\documentclass[12pt]{article}
\usepackage{graphicx}
\newcommand{\BEQ}{\begin{eqnarray} \nonumber \\}
\newcommand{\EEQ}{\\ \nonumber \end{eqnarray}}
\newcommand{\NEQ}{\\ \nonumber \\}
\newcommand{\CVECW}[2]{\left(
					   \begin{array}{c}
						   #1 \\ #2
					   \end{array}
					   \right)}
\newcommand{\MATWW}[4]{\left[
					   \begin{array}{cc}
						   #1 & #2 \\ 
						   #3 & #4
					   \end{array}
					   \right]}

\newcommand{\FRAC}[2]{\frac{\displaystyle #1}{\displaystyle #2}}
\newcommand{\ODIF}[2]{\FRAC{{\rm d} #1}{{\rm d} #2}}
\newcommand{\PDIF}[2]{\FRAC{\partial #1}{\partial #2}}
\newcommand{\DD}{{\rm d}}
\newcommand{\INTS}[1]{\int \DD #1}
\newcommand{\INTW}[2]{\int \int \DD #1 \DD #2}
\newcommand{\EE}{{\rm e}}

\newcommand{\ETAL}{{\em et al.}$\;$}
\newcommand{\OTHER}{{\rm otherwise}}
\newcommand{\HALF}{\FRAC{1}{2}}
\newcommand{\ORDER}{\mathcal{O}}
\newcommand{\AVE}[1]{\overline{#1}}
\newcommand{\SOLID}{--------}
\newcommand{\DASHED}{-\,-\,-}
\newcommand{\DASHDOT}{-\cdot-\cdot-}
\newcommand{\AAA}{a}
\newcommand{\BBB}{b}
\newcommand{\DISAB}{\rho(\AAA,\BBB)}
\newcommand{\XX}{x^{\prime}}
\newcommand{\YY}{y^{\prime}}
\newcommand{\EIGEN}{\lambda}
\newcommand{\KSIM}{\eta}
\newcommand{\SFUN}{\Phi}
\newcommand{\XSIM}{\xi}
\newcommand{\SDIS}{\Psi}
\newcommand{\RATE}{\gamma}
\newcommand{\POW}{\sigma}
\newcommand{\VAR}{s}
\newcommand{\DEL}{\Delta}
\begin{document}
%
%
\title{Effects of Randomness on Power Law Tails in Multiplicatively Interacting Stochastic Processes}
\author{Toshiya Ohtsuki, Akihiro Fujihara and Hiroshi Yamamoto \\
        {\it Graduate School of Integrated Science, Yokohama City Univ.,} \\
        {\it Kanazawa-ku, Yokohama 236-0027 Japan} }
\date{\parbox{0.8\textwidth}{\small
Effects of randomness on non-integer power law tails in multiplicatively interacting stochastic processes are investigated theoretically.   Generally, randomness causes decrease of the exponent of tails and the growth rate of processes.   Explicit calculations are performed for two examples: uniformly distributed and two peaked systems.   Significant influence is demonstrated when a bare growth rate is low and coupling is weak.   It should be emphasized that even the sign of the growth rate can be changed from positive to negative growth.
}}
\maketitle
%
Power laws are widely observed not only in natural systems but also in social phenomena [\ref{ref: Stanley}-\ref{ref: AlBa}].   A number of efforts have been made to explain these behavior [\ref{ref: BTW}-\ref{ref: Sornette}] but the underlying physics has not been clarified yet.   Recently, inelastic Maxwell models where randomly chosen particles collide inelastically with each other have been studied extensively in the context of granular materials [\ref{ref: BCG}-\ref{ref: BeKr}].   Then it has been shown that probability distribution functions of particle velocities have a non-integer power law tail asymptotically.   The exponent of tails is determined from a transcendental equation and, generally, is a continuously varying function of a parameter, i.e., a coefficient of restitution.   The system exhibits a kind of self-organized criticality.   Non-integer power laws emerge not at a special (critical) value of a parameter but in a wide range of a parameter.   ben-Avraham \ETAL [\ref{ref: ABLR}] have extended these studies to multiplicatively interacting stochastic processes.   Here the system is composed of $N (\gg 1)$ particles with positive variables $x_i>0 (i=1,2,\cdots,N)$ and evolves with multiplicative interactions between randomly chosen particles $i$ and $j$ 
\BEQ
\CVECW{x_i^{\prime}}{x_j^{\prime}} = A \CVECW{x_i}{x_j} , \hspace{10mm}
\left( A = \MATWW{\AAA}{\BBB}{\BBB}{\AAA} \right)
   \label{eq: multi-int}
\EEQ
where $\AAA$ and $\BBB$ are positive parameters.   When the eigenvalue $\lambda_A=\AAA+\BBB$ of the matrix $A$ is more (less) than unity, the system grows (declines) totally.   The case $\AAA+\BBB=1$ corresponds to inelastic Maxwell models where the total amount of variables (momenta) is conserved.   Power law tails appear at $\AAA + \BBB < 1$, $\AAA > 1$, or $\BBB > 1$ and the exponent varies continuously with $\AAA$ and $\BBB$.
\par
In actual processes, interactions sometimes accompany randomness.   Especially, mesoscopic and/or macroscopic systems such as granular meterials are composed of not identical but polydisperse particles.   In social phenomena, moreover, individuality of agents inevitably results in randomness in interactions.   Note that the process (\ref{eq: multi-int}) is a reasonable model describing wealth distribution in economical systems [\ref{ref: Slanina}].   In this Letter, therefore, we examine effects of randomness on power law tails in multiplicatively interacting systems.
\par
%
We consider processes where $\AAA$ and $\BBB$ are random parameters with probability distribution $\DISAB$.   A master equation for probability distribution functions $f(x,t)$ is given by
\BEQ
\PDIF{f}{t} + f = \INTW{\AAA}{\BBB}\DISAB \hspace{90mm} 
\nonumber \\ \nonumber \\ 
   \times \left[ \INTW{\XX}{\YY} \HALF f(\XX,t)f(\YY,t) 
   \{\delta(x-\AAA\XX-\BBB\YY)+\delta(x-\BBB\XX-\AAA\YY)\} \right] .
      \label{eq: Master eq}
\EEQ
The Fourier transform of eq.(\ref{eq: Master eq}) is
\BEQ
\PDIF{g(k,t)}{t} + g(k,t) = \INTW{\AAA}{\BBB}\DISAB  
                            g(\AAA k,t)g(\BBB k,t) ,
   \label{eq: FT eq}
\EEQ
where $g(k,t) = \INTS{x} f(x,t)\EE^{-ikx}$.   The important property of eq.(\ref{eq: FT eq}) is that its moment equations become a closed set [\ref{ref: BCG}-\ref{ref: BeKr}].   Expanding as $g(k,t) = 1 + \sum_{n=1}^{\infty} \mu_n(t) (-ik)^n /n!$ and substituting into eq.(\ref{eq: FT eq}), we have
\BEQ
\ODIF{\mu_n}{t} - \EIGEN_n \mu_n = \INTW{\AAA}{\BBB}\DISAB 
   \sum_{\ell=1}^{n-1} 
   \FRAC{n!}{\ell! (n-\ell)!}\mu_{\ell}\mu_{n-\ell}\AAA^{\ell}\BBB^{n-\ell} 
\nonumber \\ \nonumber \\
   = \sum_{\ell=1}^{n-1} 
   \FRAC{n!}{\ell! (n-\ell)!}\mu_{\ell}\mu_{n-\ell}\AVE{\AAA^{\ell}\BBB^{n-\ell}},
   \hspace{17mm}
      \label{eq: moment eq}
\NEQ
\EIGEN_n = \INTW{\AAA}{\BBB}\DISAB (\AAA^n + \BBB^n - 1)
         = \AVE{\AAA^n} + \AVE{\BBB^n} - 1.
      \label{eq: eigenvalue}
\EEQ
Here the bar denotes average with respect to $\DISAB$.   Equation (\ref{eq: moment eq}) can be solved sequentially, which enable us to treat eq.(\ref{eq: FT eq}) analytically.   Here we pursue a similarity (scaling) solution of the type
\BEQ
g(k,t) = \SFUN(\KSIM) , \hspace{10mm} (\KSIM = k \EE^{\RATE t}) 
   \label{eq: similarity sol}
\EEQ
where $\RATE$ expresses a growth rate of the system.   Substitution of eq.(\ref{eq: similarity sol}) into eq.(\ref{eq: FT eq}) leads to
\BEQ
\RATE \KSIM \ODIF{\SFUN(\KSIM)}{\KSIM} + \SFUN(\KSIM) = 
   \INTW{\AAA}{\BBB}\DISAB \SFUN(\AAA \KSIM) \SFUN(\BBB \KSIM) .
      \label{eq: similarity eq}
\EEQ
The inverse Fourier transform of $\SFUN(\KSIM)$ gives scaled distribution functions
\BEQ
\SDIS(\XSIM) = f(x,t)\EE^{\RATE t} \hspace{10mm} (\XSIM = x \EE^{-\RATE t}) .
   \label{eq: scaled dis}
\EEQ
\par
We pay our attention to tails in the small $\KSIM$ and large $\XSIM$ limit.   This limit is classified into two cases and treated separately.   In case I, we make the following ansatz,
\BEQ
\SFUN = 1 + \sum_{\ell=1}^{n} \FRAC{C_\ell}{\ell!}(-i\KSIM)^\ell 
         + C_{\POW}(-i\KSIM)^{\POW} , \hspace{10mm} (1 \leq n < \POW < n+1)
            \label{eq: ansatz A}
\EEQ
where $\POW$ represents a non-integer power.   Substituting eq.(\ref{eq: ansatz A}) into eq.(\ref{eq: similarity eq}) and equating terms of each powers $\KSIM^{\ell}$, we get
\BEQ
\ORDER(\KSIM) & \; \cdots \; & \RATE = \EIGEN_1 ,
   \label{eq: power A1}
\NEQ
\ORDER(\KSIM^{\ell}) & \; \cdots \; & 
   (\ell\RATE - \EIGEN_{\ell})C_{\ell} = \sum_{m=1}^{\ell-1}
   \FRAC{\ell!}{m! (\ell-m)!}C_{m}C_{\ell-m}\AVE{\AAA^{m}\BBB^{\ell-m}}
      \label{eq: power A2}
\\ \nonumber & & \hspace{50mm} (\ell = 2,3,\cdots,n)
\NEQ
\ORDER(\KSIM^{\POW}) & \; \cdots \; & \RATE\POW = \EIGEN_{\POW} .
   \label{eq: power A3}
\EEQ
The growth rate $\RATE$ is given by eq.(\ref{eq: power A1}) and the exponent $\POW$ is determined from the combined equation of eqs.(\ref{eq: power A1}) and (\ref{eq: power A3})
\BEQ
(\AVE{\AAA} + \AVE{\BBB} - 1)\VAR = \AVE{\AAA^{\VAR}} + \AVE{\BBB^{\VAR}} -1 .
   \label{eq: trans eq A}
\EEQ
The transcendental equation (\ref{eq: trans eq A}) always has a trivial solution $\VAR = 1$.   When a nontrivial solution $\VAR = \VAR^{\ast}$ exists in a region $\VAR > 1$, it gives the exponent $\POW = \VAR^{\ast}$.   In case II where $\VAR^{\ast} < 1$, we make the ansatz,
\BEQ
\SFUN = 1 + C_{\POW}(-i\KSIM)^{\POW} , \hspace{10mm} ( \POW < 1)
            \label{eq: ansatz B}
\EEQ
instead of eq.(\ref{eq: ansatz A}).   In this case, we have only the $\ORDER(\KSIM^{\POW})$ term, eq.(\ref{eq: power A3}), and the transcendental equation
\BEQ
\RATE \VAR = \AVE{\AAA^{\VAR}} + \AVE{\BBB^{\VAR}} -1 .
   \label{eq: trans eq B}
\EEQ
Here we need one more condition to determine two variables $\RATE$ and $\VAR$ from one equation (\ref{eq: trans eq B}).   The answer is the selection of the minimum growth rate originally reported by ben-Avraham \ETAL [\ref{ref: ABLR}].   When the line $\RATE \VAR$ and the curve $\AVE{\AAA^{\VAR}} + \AVE{\BBB^{\VAR}} -1$ come in contact at $\VAR = \VAR^{\ast\ast}$, the growth rate $\RATE$ becomes minimum and $\VAR^{\ast\ast}$ gives the exponent $\POW$ less than unity.
\par
%
Now we study effects of randomness.   First, we compare the exponent $\POW$ and the growth rate $\RATE$ with those $\POW_0$ and $\RATE_0$ in the absence of randomness, which are determined from
\BEQ
\left\{ \begin{array}{l}
\RATE_0 = \AAA_0 + \BBB_0 - 1 ,
\vspace{2mm} \\
(\AAA_0 + \BBB_0 - 1)\VAR = \AAA_0^{\VAR} + \BBB_0^{\VAR} -1 , 
\end{array} \right.
\hspace{10mm} (\VAR > 1)
   \label{eq: trans eq A0}
\NEQ
\RATE_0 \VAR = \AAA_0^{\VAR} + \BBB_0^{\VAR} -1 ,
\hspace{30mm} (\VAR < 1)
   \label{eq: trans eq B0}
\EEQ
where $\AAA_0 = \AVE{\AAA}$ and $\BBB_0 = \AVE{\BBB}$.   When $\VAR > 1$, $\AAA^{\VAR}$ is a downwards convex function of $\VAR$ and $\AVE{\AAA^{\VAR}} > \AVE{\AAA}^{\VAR}$, while $\AAA^{\VAR}$ is upward convex and $\AVE{\AAA^{\VAR}} < \AVE{\AAA}^{\VAR}$ at $\VAR < 1$.   It follows that solutions of eqs.(\ref{eq: trans eq A}) and (\ref{eq: trans eq B}) are always smaller than those of eqs.(\ref{eq: trans eq A0}) and (\ref{eq: trans eq B0}),
\BEQ
\POW < \POW_0 .
   \label{eq: pow ran-0}
\EEQ
It becomes evident that randomess causes decrease of the exponent of tails.   At the same time, the selection of the minimum growth rate leads to possible decrease of the growth rate
\BEQ
\RATE < \RATE_0 . \hspace{10mm} (\POW < 1)
   \label{eq: rate ran-0}
\EEQ
\par
\begin{figure}
\resizebox{70mm}{!}{\includegraphics{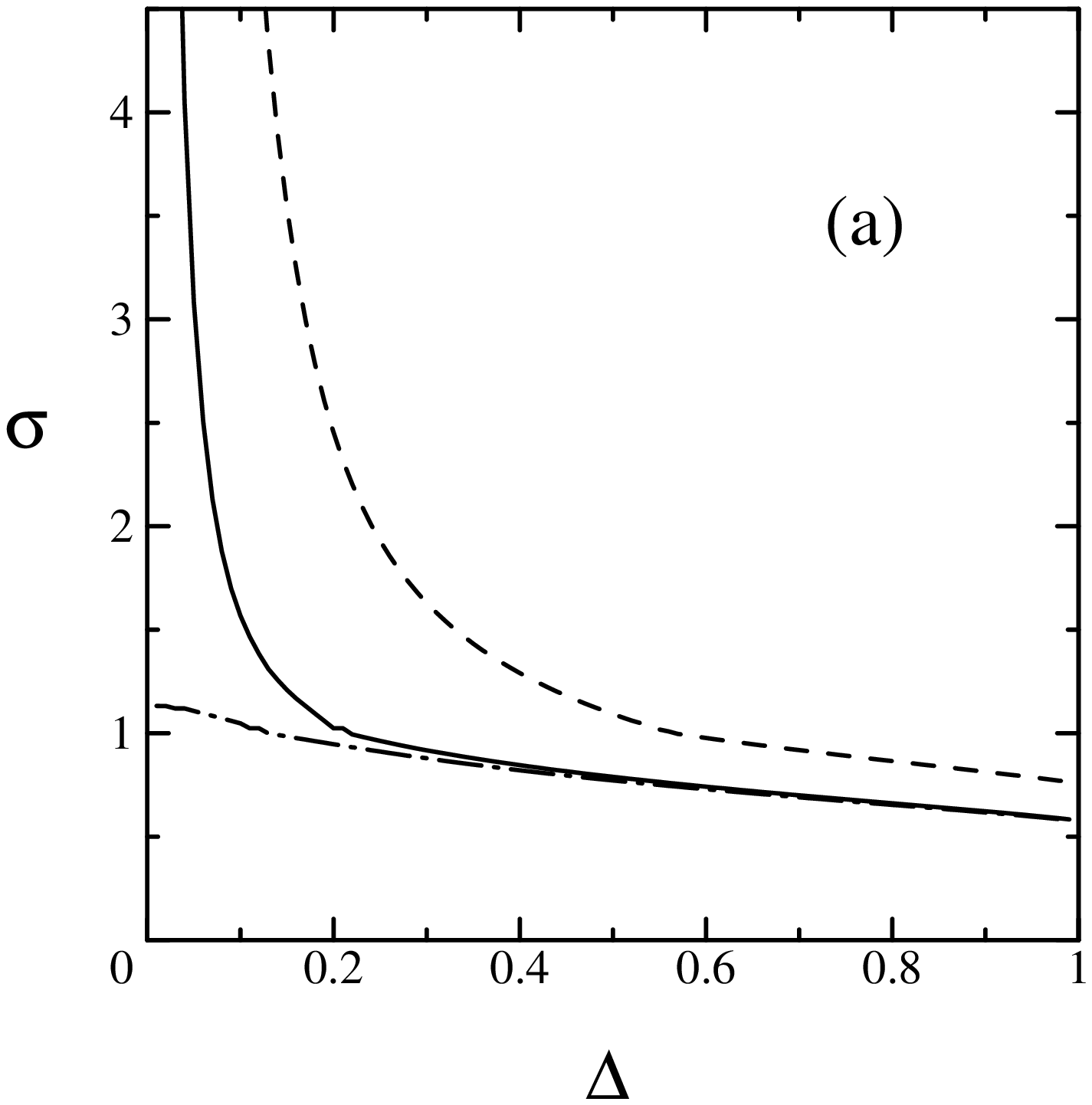}}
\hspace{10mm}
\resizebox{70mm}{!}{\includegraphics{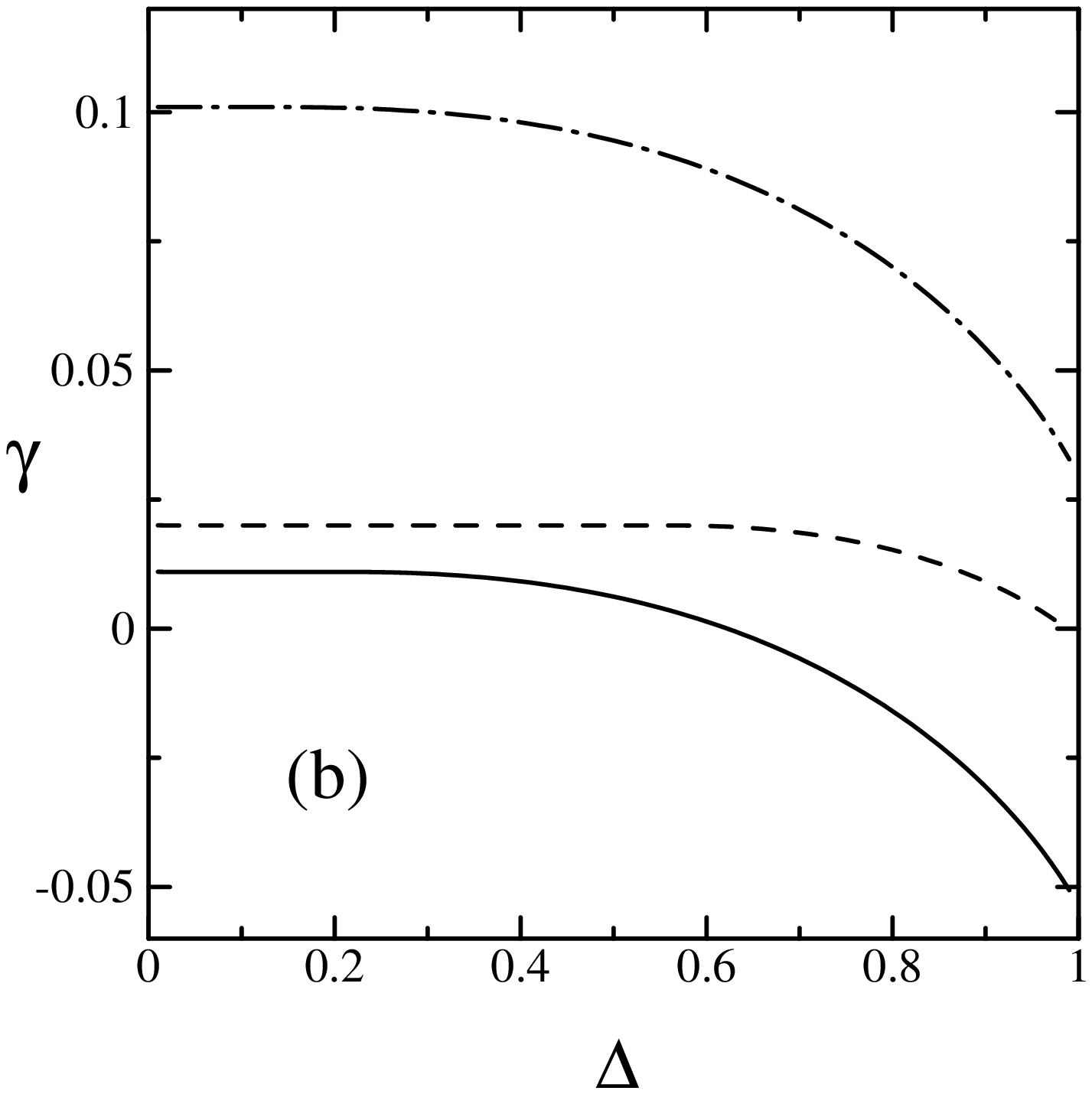}}
\caption{Calculated values of (a) exponents $\POW$ of tails and (b) growth rates $\RATE$ of processes in uniformly distributed systems at $\AAA_0 = 1.01$, $\BBB_0 = 0.001$ (\SOLID), $\AAA_0 = 1.01$, $\BBB_0 = 0.01$ ($\DASHED$) and $\AAA_0 = 1.1$, $\BBB_0 = 0.001$ ($\DASHDOT$).}
\vspace{5mm}
\label{fig: uni exp rate}
\end{figure}

Next, we carry out explicit calculations for two examples.   First example is an uniformly distributed system described by
\BEQ
\DISAB = 
\left\{ \begin{array}{ll}
\FRAC{1}{(\AAA_2-\AAA_1)(\BBB_2-\BBB_1)} , \hspace{10mm} 
   & (\AAA_1<\AAA<\AAA_2,\; \BBB_1<\BBB<\BBB_2)
\vspace{2mm} \\
0 . \hspace{10mm} & (\OTHER) 
\end{array} \right.
   \label{eq: uni dis}
\EEQ
The transcendental equations (\ref{eq: trans eq A}) and (\ref{eq: trans eq B}) read
\BEQ
(\AAA_0+\BBB_0-1)\VAR = \FRAC{1}{\VAR+1} 
   \left( \FRAC{\AAA_2^{\VAR+1}-\AAA_1^{\VAR+1}}{\AAA_2-\AAA_1}
        + \FRAC{\BBB_2^{\VAR+1}-\BBB_1^{\VAR+1}}{\BBB_2-\BBB_1} \right) - 1 ,
   \hspace{10mm} (\VAR > 1)
        \label{eq: trans uni A}
\NEQ
\RATE \VAR = \FRAC{1}{\VAR+1} 
   \left( \FRAC{\AAA_2^{\VAR+1}-\AAA_1^{\VAR+1}}{\AAA_2-\AAA_1}
        + \FRAC{\BBB_2^{\VAR+1}-\BBB_1^{\VAR+1}}{\BBB_2-\BBB_1} \right) - 1 ,
   \hspace{10mm} (\VAR < 1)
        \label{eq: trans uni B}
\EEQ
where $\AAA_0 = (\AAA_1+\AAA_2)/2$ and $\BBB_0 = (\BBB_1+\BBB_2)/2$.   Computed values of the exponent $\POW$ and the growth rate $\RATE$ are plotted in Fig.\ref{fig: uni exp rate} as a function of $\DEL \equiv (\AAA_2-\AAA_1)/(2\AAA_0) = (\BBB_2-\BBB_1)/(2\BBB_0)$.   When $\AAA_0$ is close to unity and a bare growth rate of systems is quite low, $\POW$ diminishes drastically with increasing randomness $\DEL$, whereas decrease of $\RATE$ is remarkable at small $\BBB_0$ and weak coupling.   Notice that $\RATE$ changes even its sign from positive to negative growth.
\par
\begin{figure}
\resizebox{70mm}{!}{\includegraphics{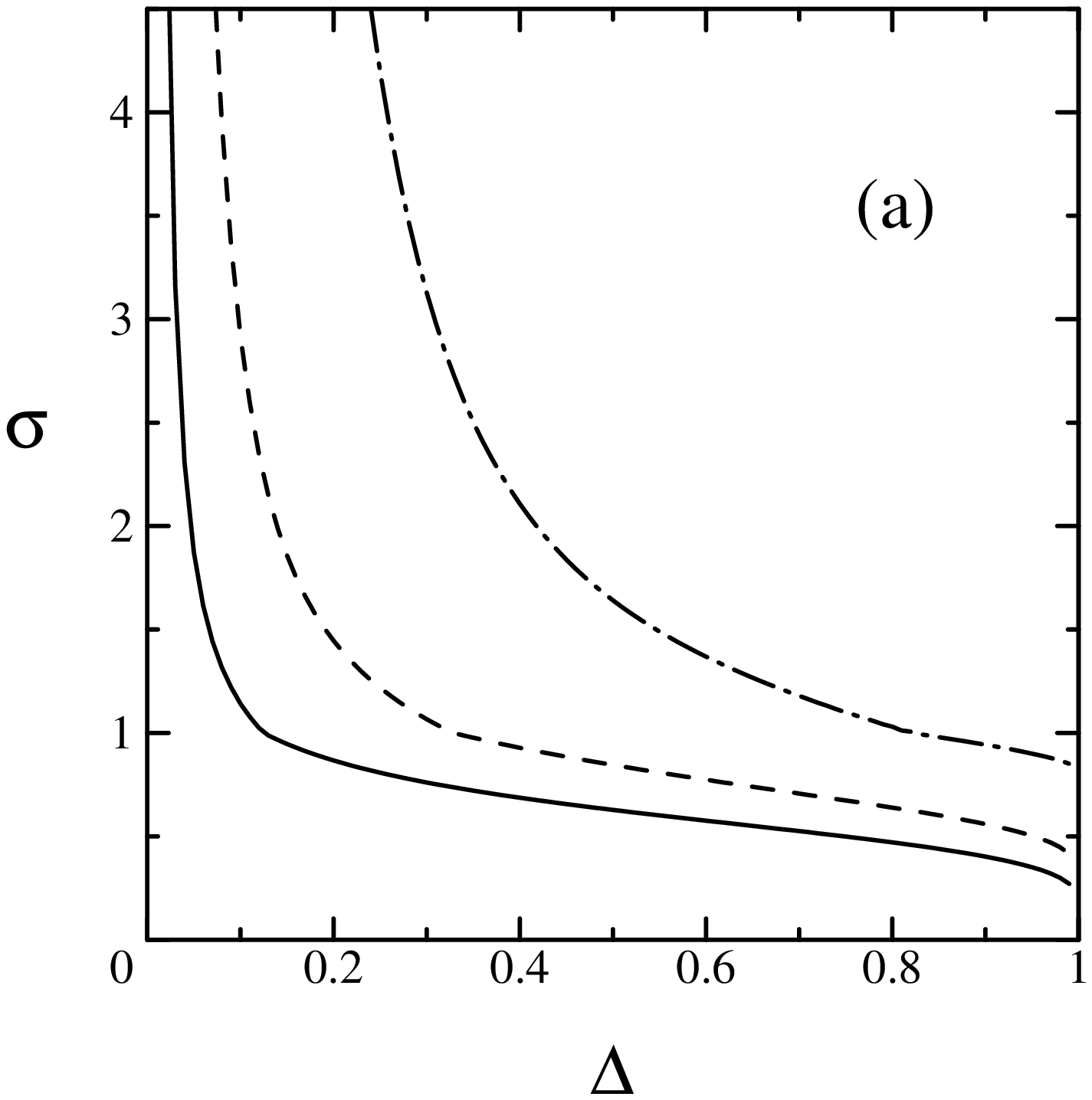}}
\hspace{10mm}
\resizebox{70mm}{!}{\includegraphics{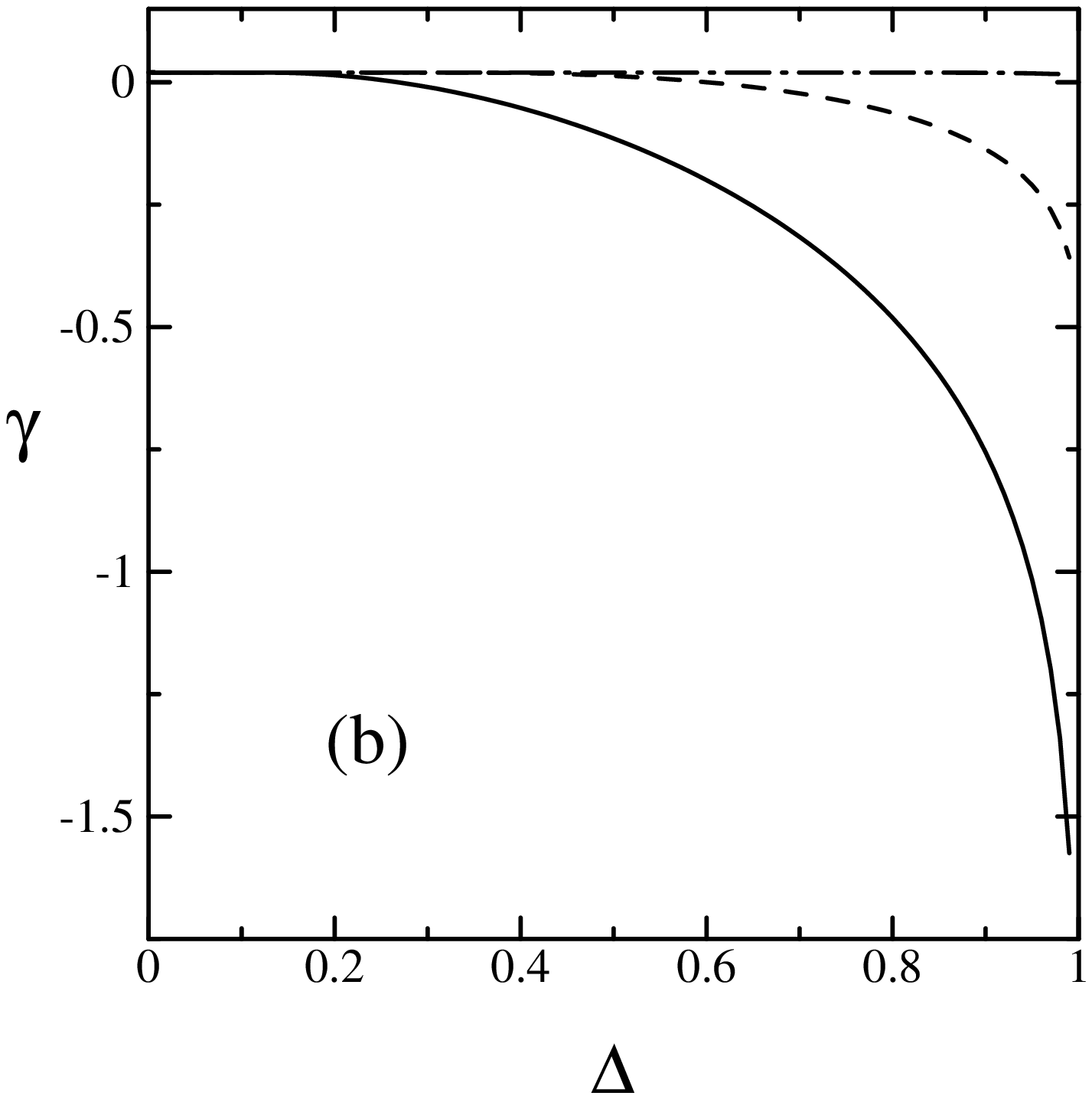}}
\caption{Calculated values of (a) exponents $\POW$ of tails and (b) growth rates $\RATE$ of processes in two peaked systems at $p=0.9$ (\SOLID), $p=0.5$ ($\DASHED$) and $p=0.1$ ($\DASHDOT$) with $\AAA_0 = 1.01$ and $\BBB_0 = 0.01$.}
\vspace{5mm}
\label{fig: 2peak exp rate}
\end{figure}

Second example is a two peaked process expressed as
\BEQ
\DISAB = p\delta(\AAA-\AAA_1)\delta(\BBB-\BBB_2)
   + (1-p)\delta(\AAA-\AAA_2)\delta(\BBB-\BBB_2) .
   \label{eq: d peak}
\EEQ
Equation (\ref{eq: d peak}) gives rise to
\BEQ
(\AAA_0+\BBB_0-1)\VAR = 
   p(\AAA_1^{\VAR}+\BBB_1^{\VAR})+(1-p)(\AAA_2^{\VAR}+\BBB_2^{\VAR}) - 1 ,
   \hspace{10mm} (\VAR > 1)
        \label{eq: trans peak A}
\NEQ
\RATE \VAR = 
   p(\AAA_1^{\VAR}+\BBB_1^{\VAR})+(1-p)(\AAA_2^{\VAR}+\BBB_2^{\VAR}) - 1 ,
   \hspace{10mm} (\VAR < 1)
        \label{eq: trans peak B}
\EEQ
where $\AAA_0 = p\AAA_1+(1-p)\AAA_2$ and $\BBB_0 = p\BBB_1+(1-p)\BBB_2$.   Figure \ref{fig: 2peak exp rate} shows numerical results, where $\DEL \equiv (1-p)(\AAA_2-\AAA_1)/\AAA_0 = (1-p)(\BBB_2-\BBB_1)/\BBB_0$.   Similarly in the case of uniform distribution, randomness $\DEL$ exerts serious influence at low bare growth rates and weak coupling.   In addition, we find that when $(1-p) \ll 1$, a few amount of impurities (aliens) with a large parameter $\AAA_2$ extremely reduce values of $\POW$ and $\RATE$.
\par   
%
From viewpoints of econophysics, conclusions obtained in this work might be important for several reasons.   (1) As mentioned before, individuality and associated randomness necessarily exist.   (2) Low bare growth rates and weak coupling are realistic conditions.   (3) The decrease of the exponent $\POW$ of tails gives enlargement of wealth differentials and has a significant economical meaning as well as the decrease of the growth rate $\RATE$.   It would be interesting to analyze wealth distribution by taking into account individuality and randomness.
%
\vspace{10mm} \\
{\Large {\bf References} }
\vspace{-5mm} \\
\renewcommand{\labelenumi}{[\arabic{enumi}]}
\begin{enumerate}
\item \label{ref: Stanley}
H. E. Stanley:
{\it Introduction to Phase Transitions and Critical Phenomena}
(Oxford, Oxford, 1971).
\item \label{ref: Mand.}
B. B. Mandelbrot:
{\it The Fractal Geometry of Nature}
(Freeman, San Francisco, 1982).
\item \label{ref: MaSt}
R. N. Mantegna and H. E. Stanley:
{\it An Introduction to Econophysics}
(Cambridge, Cambridge, 2000).
\item \label{ref: Takayasu}
H. Takayasu (Ed.):
{\it Empirical Science of Financial Fluctuations}
(Springer, Berlin, 2002).
\item \label{ref: AlBa}
R. Albert and A.-L. Barabasi:
Rev. Mod. Phys. {\bf 74} (2002) 47-97.
\item \label{ref: BTW}
P. Bak, C. Tang and K. Wiesenfeld:
Phys. Rev. Lett. {\bf 59} (1987) 381-384.
\item \label{ref: Jensen}
H. J. Jensen:
{\it Self-Organized Criticality}
(Cambridge, Cambridge, 1998).
\item \label{ref: Sornette}
D. Sornette:
{\it Critical Phenomena in Natural Sciences}
(Springer, Berlin, 2000).
\item \label{ref: BCG}
A. V. Bobylev, J. A. Carrillo and I. M. Gamba:
J. Stat. Phys. {\bf 98} (2000) 743-773.
\item \label{ref: ErBr}
M. H. Ernst and R. Brito:
J. Stat. Phys. {\bf 109} (2002) 407-432.
\item \label{ref: BeKr}
E. Ben-Naim and P. L. Krapivsky:
Phys. Rev. {\bf E66} (2002) 011309.
\item \label{ref: ABLR}
D. ben-Avraham, E. Ben-Naim, K. Lindenberg and A. Rosas:
cond-mat/0308175.
\item \label{ref: Slanina}
F. Slanina:
cond-mat/0311235.
\end{enumerate}
%
\end{document}